\documentclass[aps,prd,reprint,twocolumn,superscriptaddress,showpacs,notitlepage]{revtex4-1}

\usepackage{siunitx}
\usepackage{graphicx}
\usepackage{mathrsfs}
\usepackage{bm}
\usepackage{amsmath}
\usepackage{dcolumn}
\usepackage{epstopdf}
\usepackage{dsfont}
\usepackage{amssymb}
\usepackage{tabularx}
\usepackage{array}
\usepackage{float}
\usepackage{color}
\usepackage{epstopdf}
\usepackage{mathrsfs}
\usepackage[T1]{fontenc}
\usepackage[colorlinks, linkcolor=blue,anchorcolor=blue,citecolor=blue,urlcolor=blue]{hyperref}

\begin{document}

\title{Berry connection polarizability tensor and third-order Hall effect}

\author{Huiying Liu}
\address{Research Laboratory for Quantum Materials, Singapore University of Technology and Design, Singapore 487372, Singapore}

\author{Jianzhou Zhao}
\address{Research Laboratory for Quantum Materials, Singapore University of Technology and Design, Singapore 487372, Singapore}
\address{Co-Innovation Center for New Energetic Materials, Southwest University of Science and Technology, Mianyang 621010, China}

\author{Yue-Xin Huang}
\address{Research Laboratory for Quantum Materials, Singapore University of Technology and Design, Singapore 487372, Singapore}

\author{Xiaolong Feng}
\address{Research Laboratory for Quantum Materials, Singapore University of Technology and Design, Singapore 487372, Singapore}

\author{Cong Xiao}
\address{Department of Physics, The University of Hong Kong, Hong Kong, China}
\address{HKU-UCAS Joint Institute of Theoretical and Computational Physics at Hong Kong, China}
\address{Department of Physics, The University of Texas at Austin, Austin, Texas 78712, USA}

\author{Weikang Wu}
\address{Research Laboratory for Quantum Materials, Singapore University of Technology and Design, Singapore 487372, Singapore}
\address{Division of Physics and Applied Physics, School of Physical and Mathematical Sciences,
Nanyang Technological University, Singapore 637371, Singapore}

\author{Shen Lai}
\address{Division of Physics and Applied Physics, School of Physical and Mathematical Sciences,
Nanyang Technological University, Singapore 637371, Singapore}

\author{Wei-bo Gao}
\address{Division of Physics and Applied Physics, School of Physical and Mathematical Sciences,
Nanyang Technological University, Singapore 637371, Singapore}
\address{The Photonics Institute and Centre for Disruptive Photonic Technologies, Nanyang Technological University, Singapore 637371, Singapore}

\author{Shengyuan A. Yang}
\address{Research Laboratory for Quantum Materials, Singapore University of Technology and Design, Singapore 487372, Singapore}

\begin{abstract}
One big achievement in modern condensed matter physics is the recognition of the importance of various band geometric quantities in physical effects.
As prominent examples, Berry curvature and the Berry curvature dipole are connected to the linear and the second-order Hall effects, respectively. Here, we show that the Berry connection polarizability (BCP) tensor, as another intrinsic band geometric quantity, plays a key role in the third-order Hall effect. Based on the extended semiclassical formalism, we develop a theory for the third-order charge transport and derive explicit formulas for the third-order conductivity. Our theory is applied to the two-dimensional (2D) Dirac model to investigate the essential features of the BCP and the third-order Hall response. We further demonstrate the combination of our theory with the first-principles calculations to study a concrete material system, the monolayer FeSe. Our work establishes a foundation for the study of third-order transport effects, and reveals the third-order Hall effect as a tool for characterizing a large class of materials and for probing the BCP in band structure.

\end{abstract}

\maketitle

\section{\label{s1}Introduction}

Geometric quantities of electronic band structures play important roles in various physical properties of solids. A well known example is the Berry curvature~\cite{berry1984,xiao2010}. For the $n$th band with eigenstate $|u_n(\bm k)\rangle$,
the Berry curvature is defined as
\begin{equation}
\boldsymbol{\Omega}(\bm k)=\nabla_{\boldsymbol{k}}\times\boldsymbol{\mathcal{A}}(\bm k),
\end{equation}
where $\boldsymbol{\mathcal{A}}=\langle u_{n}|i\nabla_{\boldsymbol{k}}|u_{n}\rangle$ is the Berry connection. The Berry connection $\boldsymbol{\mathcal{A}}$ and the Berry curvature $\boldsymbol{\Omega}$ resemble the vector potential and the magnetic field in the reciprocal space. Indeed, in the semiclassical equation of motion, the Berry curvature leads to an anomalous velocity term $\sim -\dot{\bm{k}}\times \boldsymbol{\Omega}$, analogous to the Lorentz force~\cite{chang1995,sundaram1999,xiao2010}. In magnetic materials, this term typically leads to a Hall current transverse to the applied electric field, known as the intrinsic contribution to the (linear) anomalous Hall effect~\cite{onoda2002,jungwirth2002,nagaosa2010}.

In nonmagnetic materials, the linear Hall response is forbidden by the time-reversal symmetry. (Throughout this work, we consider transport in the absence of magnetic field.) Nevertheless, when the inversion symmetry is broken, there could be a second-order Hall response~\cite{sodemann2015,low2015,zhang2018,ma2019,kang2019,du2019,xiao2019,konig2019,nandy2019,zhou2020}. Sodemann and Fu~\cite{sodemann2015} showed that one contribution to this effect can be attributed to the dipole moment of Berry curvature over the occupied states.

Now, consider a nonmagnetic material with inversion symmetry or a twofold rotation in the transport plane, which actually belongs to a large class of existing materials.  Both its first- and second-order Hall responses will be suppressed by the time-reversal and the crystal symmetries, so the leading Hall response will be of the third order. Indeed, the phenomenon was observed in our recent experimental work on bulk MoTe$_2$ samples~\cite{lai2021}. Then, a natural question is,  What is the geometric quantity involved in this third-order Hall effect?

In this work, we address this question by showing that the third-order Hall effect (THE) is intimately connected to the Berry connection polarizability (BCP). The BCP is a band geometric quantity proposed by Gao \emph{et al.}~\cite{gao2014,gao2015} in formulating a generalized semiclassical theory. By definition, it is a second-rank tensor representing the ratio between the field-induced Berry connection ${\mathcal{A}}^{(1)}$ and the applied electric field~\cite{gao2014}, i.e.,
\begin{equation}
  G_{ab}(\bm k)=\frac{\partial {\mathcal{A}}^{(1)}_a(\bm k)}{\partial E_b},
\end{equation}
where the letters $a, b, \cdots$ are used to denote Cartesian coordinates. Note that although the Berry connection is gauge dependent, its polarizability is a gauge-invariant quantity, and the importance of the BCP in magnetotransport effects has been discussed before~\cite{gao2014,gao2017}.
Here, we extend the discussion in Ref.~\cite{lai2021} and develop a theory for the THE based on the generalized semiclassical theory framework. We show that the BCP is the key geometric quantity involved in the THE response linear in the relaxation time. We analyze the symmetry constraints on the BCP and the THE. We apply our theory to a two-dimensional (2D) Dirac model, revealing key features of the BCP tensor. Furthermore, combined with first-principles calculations, we demonstrate the calculations of the BCP and the third-order conductivity tensor for a real material FeSe. Our work provides the theoretical understanding for the THE, which can be used for characterizing a large class of materials not accessible with first- and second-order Hall responses. It also reveals the physical significance of the BCP and suggests the THE as a tool for probing this intrinsic band property.

\section{Third-order current response}

The third-order current response $j^{(3)}$ to an applied electric field is characterized by the third-order conductivity tensor $\chi$, with
\begin{equation}\label{chi}
  j_a^{(3)}=\chi_{abcd}E_bE_cE_d,
\end{equation}
where we have adopted the Einstein summation convention. In practice, the third-order effect can be measured by applying an AC driving with frequency $\omega$, and extracting the response signal at $3\omega$. In a typical experimental setup, the sample has a planar geometry, and the transport plane, i.e., the plane formed by the applied $E$ field and the measured response current, coincides with the sample plane (usually denoted as the $x$-$y$ plane). In such a case, the subscripts $a,b,c,d\in \{x,y\}$.

Clearly, the third-order response has symmetry properties distinct from the second order. For the second-order response with $j^{(2)}_a=\chi_{abc}E_b E_c$, since both the current and the $E$ field are polar vectors, the response vanishes as long as the transport plane has an inversion center. It should be noted that this includes but is not limited to crystals with inversion symmetry. For example, a twofold (screw) rotation along $z$ can also suppress the second-order response in the $x$-$y$ plane. In such cases, given that the linear Hall response is suppressed by the time-reversal symmetry, the third-order response will dominate the Hall transport. From this discussion, one can see that the THE actually dominates in a large class of materials.

We develop a theory for the THE within the generalized semiclassical theory~\cite{gao2014,gao2015,gao2019}. In the semiclassical framework, the total current response for a uniform system can be expressed as (we set $e=\hbar=1$)
\begin{equation}
\boldsymbol{j}=-\int[d\boldsymbol{k}]\mathcal{D}(\bm k)\dot{\boldsymbol{r}}f(\boldsymbol{k}),\label{eq:currentdef}
\end{equation}
where $[d\boldsymbol{k}]$ is a shorthand notation for $\sum_n d\bm k/(2\pi)^d$ with $d$ the dimension of the system, {and the quantities inside the integral are understood to carry an implicit band index $n$.} {$\mathcal{D}(\bm k)$ is the correction factor for the phase space density of states~\cite{xiao2005,xiao2010},} and here we have $\mathcal{D}(\bm k)=1$ in the absence of magnetic field. $f$ is the single-particle distribution function.

With the time-reversal symmetry, the third-order conductivity tensor must start from terms linear in the relaxation time $\tau$. There is no ``intrinsic'' contribution from the equilibrium distribution $f_0$. In other words, the THE must be derived from the nonequilibrium distribution $\delta f=f-f_0$. Since the nonequilibrium distribution is at least of linear order in the $E$ field, to obtain the third-order current $j \sim E^3$, the electron velocity $\dot{\bm r}$ must be accurate to the second order in $E$.

In the original semiclassical theory formulated by Chang, Sundaram, and Niu~\cite{chang1995,sundaram1999}, the equations of motion are only accurate to the first order in external fields, which is insufficient for our task. Fortunately, in Refs.~\cite{gao2014,gao2015}, Gao, Yang, and Niu extended the theory to higher order. The semiclassical equations of motion with second-order accuracy are given by
\begin{gather}
\dot{\boldsymbol{r}}=\frac{\partial\tilde{\varepsilon}}{\partial\boldsymbol{k}}-\dot{\boldsymbol{k}}\times\tilde{\boldsymbol{\Omega}},\label{eq:EOD1}\\
\dot{\boldsymbol{k}}=-\boldsymbol{E}-\dot{\bm r}\times \bm B.
\end{gather}
Differing from the Chang-Sundaram-Niu equations, $\tilde{\varepsilon}$ and $\tilde{\boldsymbol{\Omega}}$ are the energy and the Berry curvature including field corrections. In the absence of $B$ field, $\tilde{\varepsilon}$ accurate to the second order in $E$ field is given by (here we add the band index $n$)
\begin{equation}\label{Ecorr}
\tilde{\varepsilon}_{n}=\varepsilon_{n}-\sum_{m\neq n}\frac{(\boldsymbol{E}\cdot\boldsymbol{\mathcal{A}}_{nm})(\boldsymbol{E}\cdot\boldsymbol{\mathcal{A}}_{mn})}{\varepsilon_{n}-\varepsilon_{m}},
\end{equation}
where $\varepsilon_{n}$ is the unperturbed band energy, and $\boldsymbol{\mathcal{A}}_{nm}=\langle u_n|i\nabla_{\bm k}|u_m\rangle$ is the interband Berry connection.  Note that $\dot{\bm k}$ is already of first order in $E$; to have $\dot{\bm r}$ accurate to $E^2$ order, we only need the Berry curvature $\tilde{\boldsymbol{\Omega}}$ corrected to first order in $E$:
\begin{equation}\label{Omegap}
  \tilde{\bm{\Omega}}(\bm k)=\bm{\Omega}(\bm k)+\bm{\Omega}^{(1)}(\bm k).
\end{equation}
The first-order correction to the Berry curvature
\begin{equation}\label{Bind}
  \bm{\Omega}^{(1)}=\nabla_{\bm k}\times \bm{\mathcal{A}}^{(1)},
\end{equation}
which naturally involves the BCP, as
\begin{equation}\label{Ap}
  \mathcal{A}_a^{(1)}(\bm k)=G_{ab}(\bm k) E_b.
\end{equation}
As an intrinsic band geometric quantity, the BCP can be expressed solely in terms of the (unperturbed) band eigenstates $|u_n\rangle$ and band energies $\varepsilon_n$, with
\begin{equation}
G_{ab}=2 \text{Re}\sum_{m\neq n}\frac{(\mathcal{A}_{a})_{nm}(\mathcal{A}_{b})_{mn}}{\varepsilon_{n}-\varepsilon_{m}}.\label{eq:Gtensor}
\end{equation}
From Eqs.~(\ref{Omegap}) and (\ref{Ap}), we may also define a Berry curvature polarizability as
\begin{equation}\label{Ptensor}
  P_{ab}=\frac{\partial \tilde{\Omega}_a}{\partial E_b}=\epsilon_{acd}\partial_c G_{db},
\end{equation}
where $\epsilon$ is the Levi-Civita antisymmetric tensor, and $\partial_c$ is a shorthand notation for $\partial/\partial k_c$.
Interestingly, from Eqs.~(\ref{eq:Gtensor}) and (\ref{Ecorr}), one notes that the second-order energy correction $\varepsilon^{(2)}$, i.e., the last term in (\ref{Ecorr}), can also be expressed with the BCP tensor as
\begin{equation}
  \varepsilon^{(2)}=-\frac{1}{2}E_a G_{ab} E_b.
\end{equation}
This will be used in the following derivation.

Meanwhile, the distribution function $f$ in Eq.~(\ref{eq:currentdef}) is solved from the Boltzmann equation within the relaxation time approximation:
\begin{equation}
\dot{\boldsymbol{k}}\cdot\nabla_{\boldsymbol{k}}f=-\frac{f-f_0}{\tau},
\end{equation}
where $f_0$ is the equilibrium Fermi-Dirac distribution, and $\tau$ is the transport relaxation time. Combined with the semiclassical equations of motion, the distribution function is obtained as a series expansion
\begin{equation}
f=\sum_{\alpha=0}^\infty(\tau\boldsymbol{E}\cdot\nabla_{\boldsymbol{k}})^{\alpha}f_{0}(\tilde{\varepsilon}).
\end{equation}

Now, we substitute the expressions for $\dot{\bm r}$ and $f$ into the current $\bm j$ in Eq.~(\ref{eq:currentdef}), and collect terms $\sim E^3$.  The obtained third-order current contains the following terms:
\begin{widetext}
 \begin{equation}\begin{split}\label{j3}
  \bm j^{(3)}=&-\tau\int [d\bm k]\nabla_{\bm k}\varepsilon\ (\bm E\cdot\nabla_{\bm k}) [\varepsilon^{(2)} f_0']-\tau\int [d\bm k]\nabla_{\bm k}\varepsilon^{(2)}\ (\bm E\cdot\nabla_{\bm k}) f_0 -\tau\int[d\bm k] \bm E\times\bm{\Omega}^{(1)}\ (\bm E\cdot\nabla_{\bm k})f_0\\
  &\quad -\tau^3\int [d\bm k]\nabla_{\bm k}\varepsilon\ (\bm E\cdot\nabla_{\bm k})^3 f_0.
  \end{split}
\end{equation}
\end{widetext}
Here, the first term comes from the second-order energy correction in the distribution function, the second term is from the second-order correction to the velocity, the third term is due to the anomalous velocity from the field-induced Berry curvature, and the last term is from the $E^3$ term in the nonequilibrium distribution. In terms of the relaxation time, $j^{(3)}$ contains only terms with odd powers of $\tau$, consistent with the requirement of time-reversal symmetry of the system. In addition, all terms contain the derivatives of the Fermi distribution, demonstrating that the transport current is from the Fermi surface, as it should be.

Comparing Eq.~(\ref{j3}) with Eq.~(\ref{chi}), we obtain the result for the third-order conductivity tensor. It contains two parts: one part $\chi^\text{I}$ is linear in $\tau$, and the other part $\chi^\text{II}$ is proportional to $\tau^3$. Explicitly, we have
\begin{equation}\begin{split}\label{chiI}
  \chi^\text{I}_{abcd}=&\ \tau\int [d\bm k] (-\partial_a \partial_b G_{cd}+\partial_a\partial_d G_{bc}-\partial_b\partial_d G_{ac})f_0\\
  &+\frac{\tau}{2}\int [d\bm k]v_a v_b G_{cd} f_0'',
  \end{split}
\end{equation}
and
\begin{equation}
  \chi^\text{II}_{abcd}=-\tau^3\int [d\bm k] v_a\partial_b\partial_c\partial_d f_0,
\end{equation}
where $\bm v=\nabla_{\bm k}\varepsilon$ is the (unperturbed) band velocity. {For the first term in Eq.~(\ref{chiI}), we have put the derivatives on $G$ via integration by parts.} 
Here, $\chi^\text{I}$ and $\chi^\text{II}$ represent distinct contributions, as they have distinct scaling in $\tau$.
In experiment, this distinction enables the separation of the two when plotted versus the (linear) longitudinal conductivity $\sigma_{xx}$.
In the following, we shall focus on the $\tau$-linear contribution  $\chi^\text{I}$. This is because first,
$\chi^\text{I}$ is connected with the BCP which is of our interest (in comparison, $\chi^\text{II}$ is only related to the band dispersion), and second,
$\chi^\text{I}$ can be more reliably extracted in experiment through the scaling analysis~\cite{lai2021}.

 It should be noted that the obtained conductivity tensor contains information for both longitudinal and transverse third-order current response.
If we consider the power dissipation $\sim j_a E_a=\chi_{abcd}E_a E_b E_c E_d$, we may decompose the nonlinear conductivity into two parts
$\chi_{abcd}=\chi_{(ab)cd}+\chi_{[ab]cd}$, where $\chi_{(ab)cd}$ ($\chi_{[ab]cd}$) is the symmetric (antisymmetric) part of the tensor with respect to the first two indices (after symmetrizing the last three indices). Clearly, the symmetric part contributes to the dissipation, whereas the antisymmetric part does not. In analogy with the linear Hall effect which is dissipationless, we may choose to define $\chi_{[ab]cd}$ as the Hall part for the third-order response. Nevertheless, in experiment, what is directly measured is the current response transverse to the applied $E$ field, so in the following discussion, we will focus on the quantity that directly captures this response in order to facilitate the comparison with experiment.

\section{2D Dirac model}

In terms of symmetry, the BCP is a second-rank symmetric polar $i$-tensor, whereas the third-order conductivity $\chi$ is a fourth-rank polar $c$-tensor. {$i$-tensor ($c$-tensor) means its component is invariant (changes sign) under time-reversal operation.}  To investigate their features, we apply the above theory to analyze the 2D Dirac model. Despite its simplicity, the model describes the essential physics for a range of physical systems, including many 2D materials such as graphene~\cite{castro2009} as well as the surface of topological insulators~\cite{hasan2010,qi2011}.

The model to be considered can be written as
\begin{equation}\label{Dirac}
H(\boldsymbol{k})=w k_{x}\sigma_{0}+v_{x}k_{x}\sigma_{x}+v_{y}k_{y}\sigma_{y}+\Delta \sigma_{z},
\end{equation}
where the $\sigma$'s are the Pauli matrices, $\sigma_0$ is the $2\times 2$ identity matrix, and $w$, $v_{x}$, $v_{y}$, and $\Delta$ are model parameters. The last term gives a gap of $2\Delta$ of the Dirac spectrum, and $\Delta=0$ leads to a Dirac point. Here, we also include a tilt term, i.e., the first term in (\ref{Dirac}), which typically exists when the Dirac point is off the high-symmetry point.

\begin{figure}[tb!]
\centering
\includegraphics[width=1\columnwidth]{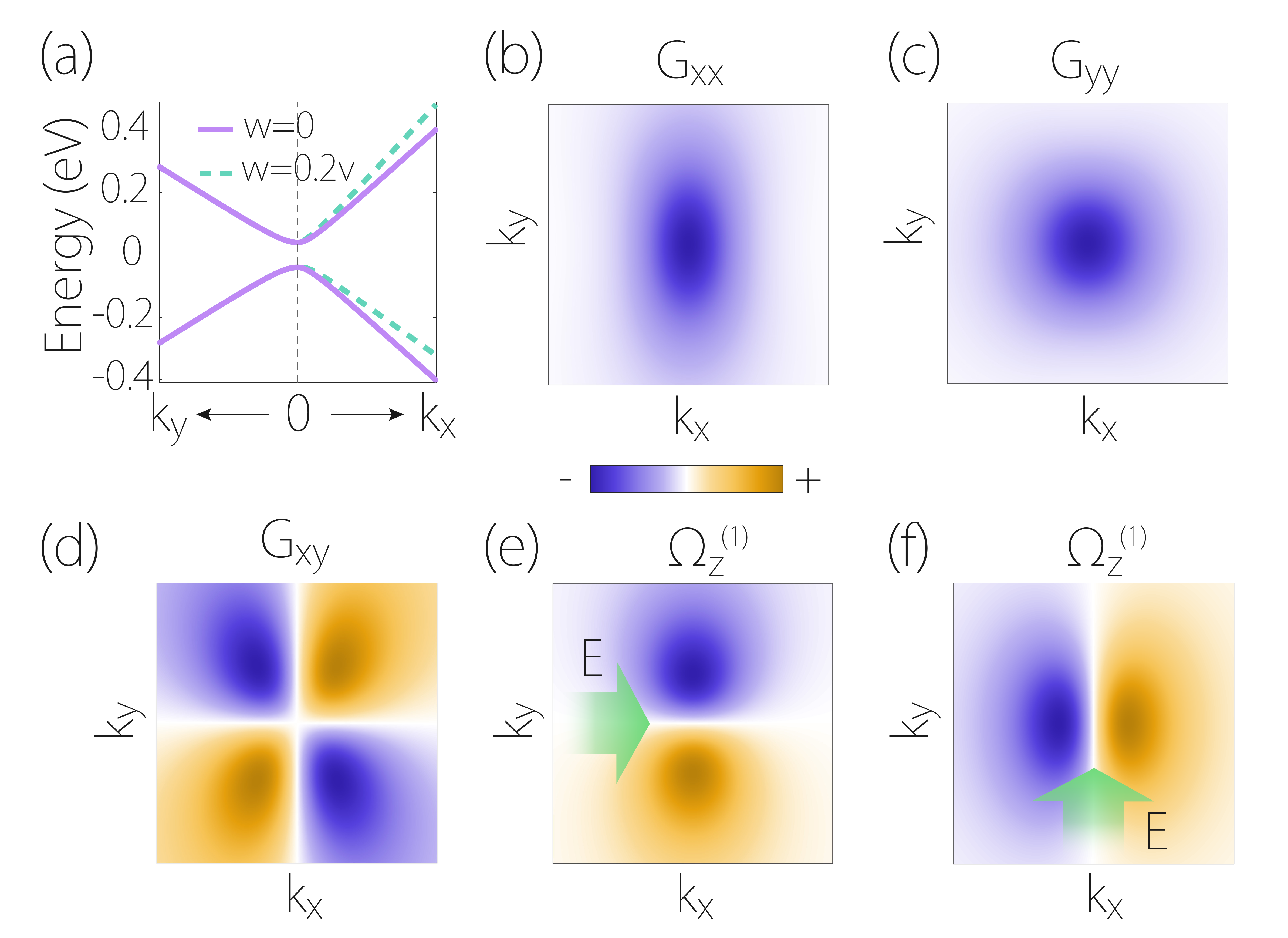}
\caption{(a) Energy spectrum of the 2D Dirac model along $k_{x}$ and $k_{y}$ axes. The solid (dashed) lines are for the case without (with) the tilt term. (b-d) Distribution of BCP tensor components for the valence band. (e,f) show the field-induced Berry curvature $\Omega_{z}^{(1)}$ when the electric field (denoted by the green arrow) is along (e) $x$ and (f) $y$. In the calculation, we take $v_{x}=1\times10^{6}$ m/s, $v_{y}=0.7v_x$, $w=0.2 v_x$, and $\Delta=40$ meV. }
\label{fig1}
\end{figure}

The model contains two bands with the following Dirac-type dispersion,
\begin{equation}
\varepsilon_{\pm}(\bm k)=w k_{x}\pm\sqrt{v_{x}^{2}k_{x}^{2}+v_{y}^{2}k_{y}^{2}+\Delta^{2}},
\end{equation}
where $\pm$ denotes the conduction and the valence band, respectively. This band structure is shown in Fig.~\ref{fig1}(a). The BCP tensor can be directly evaluated from Eq.~(\ref{eq:Gtensor}). For this simple model, we obtain for the valance band
\begin{equation}
[G_{ab}]=-\frac{v_x^2 v_y^2}{4\Lambda^{5}}\begin{bmatrix}k_{y}^{2}+\Delta^{2}/v_{y}^{2} & -k_{x}k_{y}\\
 -k_{x}k_{y} & k_{x}^{2}+\Delta^{2}/v_{x}^{2}
\end{bmatrix},
\end{equation}
where $\Lambda=\sqrt{v_{x}^{2}k_{x}^{2}+v_{y}^{2}k_{y}^{2}+\Delta^{2}}$ is the energy splitting between the two bands. One notes that the BCP does not depend on the tilt term. This is because the tilt term is an overall $k$-dependent energy shift for both bands, and according to Eq.~(\ref{eq:Gtensor}), the BCP is not affected by such overall shifts. The components of the BCP tensor are plotted in Fig.~\ref{fig1}(b-d). One observes that the diagonal components of the BCP have a monopole-like structure, which is peaked at the center where the gap is minimal. In comparison, the off-diagonal component $G_{xy}$ exhibits a quadrupole-like structure [see Fig.~\ref{fig1}(d)]. From Fig.~\ref{fig1} and the formula for the BCP, one can see that like the Berry curvature, the BCP is generally peaked around small band gaps.

Under an in-plane electric field $\bm E$, there will be a field-induced Berry curvature $\bm{\Omega}^{(1)}$. According to Eqs.~(\ref{Bind})-(\ref{Ptensor}), we find that
\begin{equation}
\bm{\Omega}^{(1)}(\boldsymbol{k})=\frac{v_{x}^{2}v_{y}^{2}}{2\Lambda^{5}}\bm k\times\bm E.
\end{equation}
One notes that although the direction of $\bm{\Omega}^{(1)}$ is constrained to be out-of-plane due to the 2D character, the induced Berry curvature $\Omega^{(1)}\sim k$, exhibiting a dipole like structure, as shown in Fig.~\ref{fig1}(e,f). This is in contrast to the original Berry curvature, which takes a monopole structure. In addition, the dipole's direction depends on the applied $E$ field, and it can be rotated by rotating the $E$ field, as illustrated in Fig.~\ref{fig1}(e,f).

\begin{figure}[tb!]
\centering
\includegraphics[width=1\columnwidth]{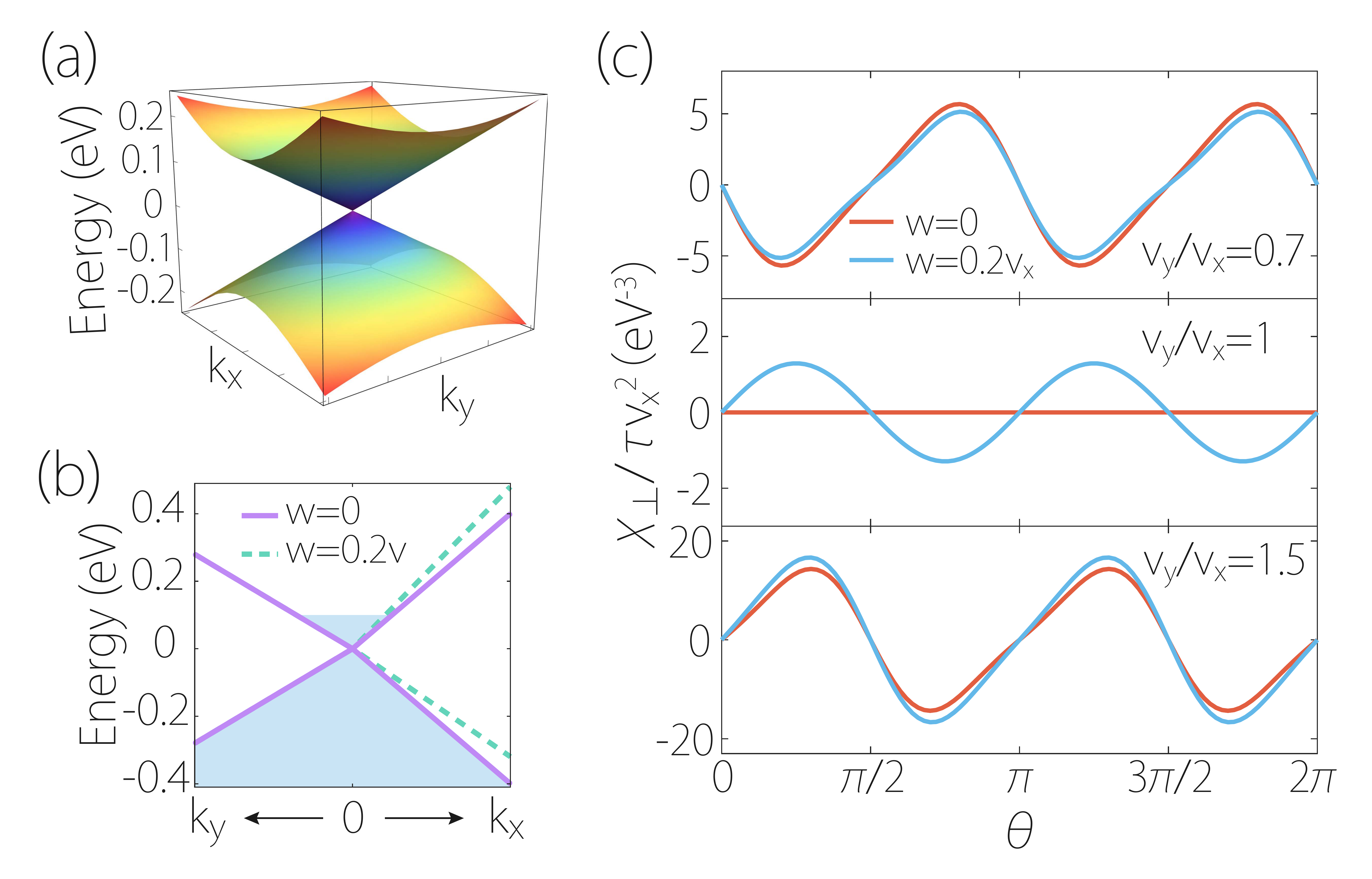}
\caption{Band structure  of the 2D Dirac model with $\Delta=0$. (b) shows the spectrum along the $k_x$ and $k_y$ axes. The solid (dashed) lines are for the case without (with) the tilt term. (c) Calculated third-order transverse conductivity $\chi_\bot$ versus the angle $\theta$ ($\theta$ is the angle between the $E$ field and the $x$ direction).
In (b), we take $v_{x}=1\times10^{6}$ m/s and  $v_{y}=0.7v_x$. In (c), we take $v_{x}=1\times10^{6}$ m/s and chemical potential $\mu=0.1$ eV.}
\label{fig2}
\end{figure}

With the knowledge of the BCP tensor, we evaluate the third-order conductivity tensor according to Eq.~(\ref{chiI}). For simplicity, we take the gap term $\Delta=0$ [see Fig.~\ref{fig2}(a,b)]. In this case, we have a Dirac point located on a mirror line with $M_y=\sigma_x$, such that $M_y H(k_x,k_y)M_y^{-1}=H(k_x,-k_y)$. The mirror symmetry dictates that the tensor elements like $\chi_{xyyy}$ and $\chi_{yxxx}$ must vanish. In other words, there is no third-order Hall response when the applied $E$ field is along or perpendicular to a mirror line. Nevertheless, the Hall response can still exist for other field orientations.

 In nonlinear transport experiments, such as in Refs.~\cite{kang2019,lai2021}, the standard practice is to fabricate multiple leads to a disk-shaped sample and do angle-resolved measurement, where the applied $E$ field rotates in the plane and one measures the current response transverse to the applied field. For such a case, it is convenient to calculate the nonlinear conductivity tensor in the fixed coordinate frame of the crystal, and then express the transverse conductivity by these tensor components and the rotation angle.

For example, consider a 2D setup, with the $E$ field along an in-plane direction specified by the polar angle $\theta$ in the fixed crystal frame. 
Then,  $\bm E=E(\cos\theta,\sin\theta,0)$, and the third-order transverse current is
\begin{equation}
   j^{(3)}_\bot(\theta)=\bm j^{(3)}\cdot (\hat{z}\times \hat{\bm E}),
\end{equation}
 which is characterized by the third-order transverse conductivity defined as
\begin{equation}
  \chi_\bot(\theta)=\frac{j^{(3)}_\bot}{E^3}.
\end{equation}
{ As we have mentioned, this $\chi_\bot(\theta)$ is what is actually measured in experiment, so it is of our primary interest. }
For a 2D system with a mirror line along $x$, as for our model with a Dirac point, we have
\begin{equation}
\begin{split}
\chi_{\bot}(\theta)=& (-\chi_{11}+3\chi_{21})\cos^{3}\theta\sin\theta\\
& \quad+(\chi_{22}-3\chi_{12})\cos\theta\sin^{3}\theta,\label{eq:chi_theta}
\end{split}
\end{equation}
where $\theta$ is measured from the mirror line, and we define the short-hand notations that $\chi_{11}=\chi_{xxxx}$, $\chi_{22}=\chi_{yyyy}$, $\chi_{12}=(\chi_{xxyy}+\chi_{xyxy}+\chi_{xyyx})/3$,
and $\chi_{21}=(\chi_{yyxx}+\chi_{yxyx}+\chi_{yxxy})/3$.



We numerically evaluate $\chi_\bot(\theta)$ for our 2D Dirac
model, and the obtained results are plotted in Fig.~\ref{fig2}(c). One
 observes that $\chi_\bot$
varies with the direction of the applied electric field
in the period of $\pi$. The response vanishes when $\theta$ equals multiples of $\pi/2$, i.e., when the transport is along or perpendicular to the mirror line, consistent with our previous analysis. Comparing the top and bottom panels of Fig.~\ref{fig2}(c), we see that for this model, the sign of $\chi_\bot$ depends on the ratio of $v_x/v_y$. For the special case with $v_x=v_y$ and $w=0$, the model becomes isotropic and third-order transverse conductivity vanishes for all angles, as shown in the middle panel of Fig.~\ref{fig2}(c).


Here, we have adopted the 2D Dirac model to demonstrate some typical features of the effect. {The obtained result may also help to understand the effect in systems with multiple Dirac points. In graphene, the effect vanishes identically, as it is suppressed by the $C_{3v}$ symmetry. Explicitly, under $C_3$, we must have $\chi_{11}=\chi_{22}=3\chi_{12}=3\chi_{21}$ in Eq.~(\ref{eq:chi_theta})~\cite{boyd2008}, so $\chi_\bot$ vanishes identically. When $C_3$ is broken, e.g., by applied strain, the effect can become nonzero. Then, the contribution from each valley can be described by our model, and the two valleys will add up.}

The symmetry properties of $\chi_\bot$ can be analyzed in the standard way. For a 2D system, we find that $C_3$ and $C_6$ symmetries would make
$\chi_\bot$ isotropic. Furthermore, the third-order transverse current is suppressed for 2D systems with point groups of $C_{3v}$, $C_{6v}$, $D_{3}$, $D_{3h}$, $D_{3d}$, $D_{6}$, or $D_{6h}$.


\section{A material example}

In the following section, we study the THE in a
concrete material, monolayer FeSe, using first-principles calculations.

\begin{figure}[tb!]
\centering
\includegraphics[width=1\columnwidth]{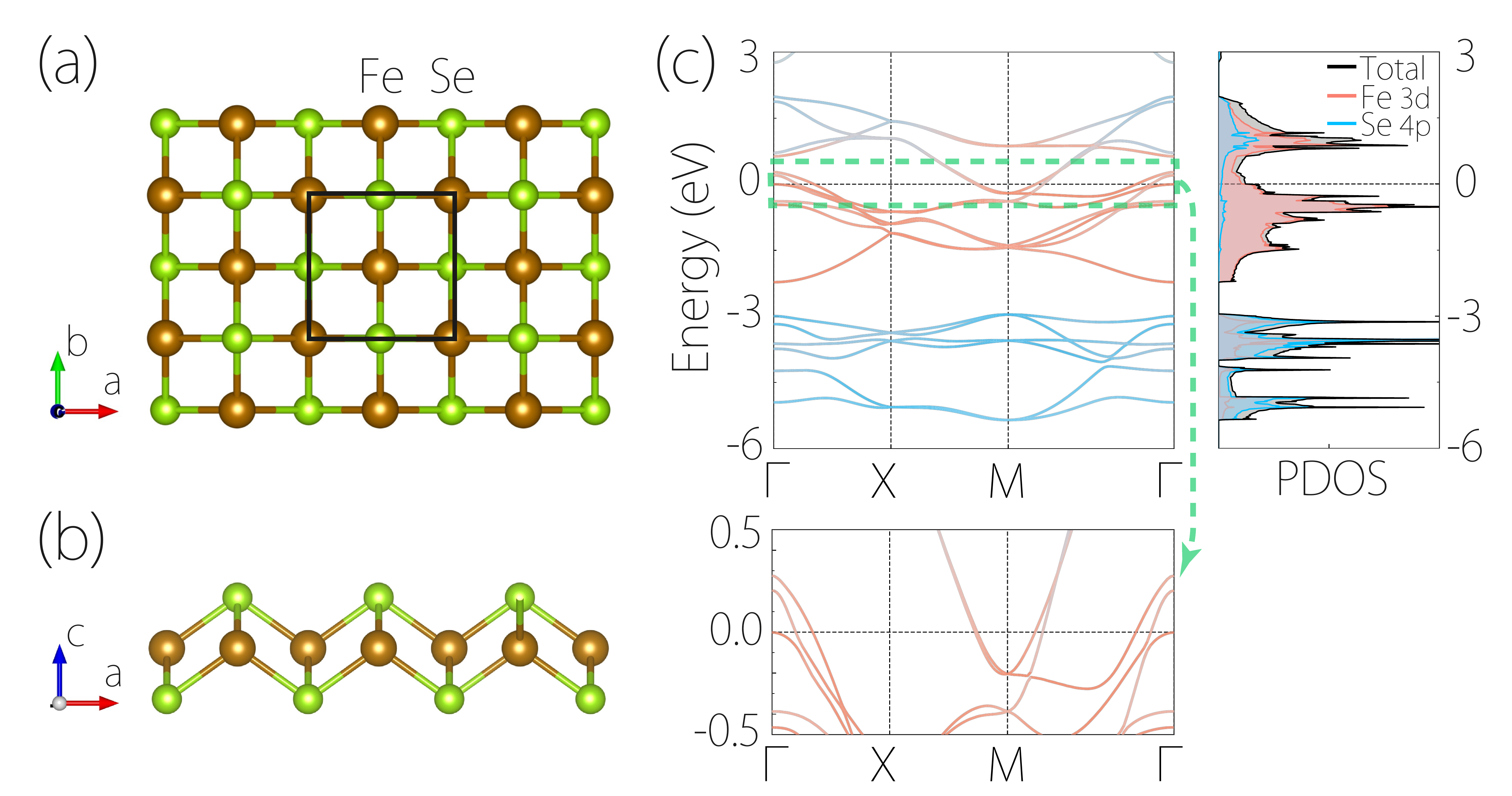}

\caption{(a) Top and (b) side view of monolayer FeSe. Brown (green) balls stand for Fe (Se) atoms. Black lines in (a) indicate the unit cell. (c) Calculated band structure and PDOS of monolayer FeSe in the nonmagnetic phase. The red (blue) color indicates the Fe-3$d$ and Se-4$p$ orbital components. The lower panel in (c) shows the enlarged view of the band structure around the Fermi level.}
\label{fig3}
\end{figure}

FeSe is a well known member of the iron-based superconductor material family~\cite{stewart2011}. It has a layered structure and its bulk superconducting $T_{c}$
is about 9 K~\cite{coldea2018}. {Bulk FeSe undergoes a tetragonal-to-orthorhombic structural transition at about 90 K~\cite{coldea2018}}.  Monolayer FeSe has been successfully fabricated, and studied on different substrates such as SiC~\cite{song2011} and SrTiO$_{3}$~\cite{wang2012,ge2015}. We shall consider monolayer FeSe in the high-temperature nonmagnetic phase. The crystal structure of monolayer FeSe is shown in Fig.~\ref{fig3}(a,b). It has a square lattice with space group $P4/nmm$ (No.~129). The system possesses the inversion symmetry, the fourfold rotation $C_{4z}$, and four mirror lines in the 2D plane along the [10], [01], [11], and [1$\bar{1}$] directions. Clearly, for this system, the linear and second Hall responses are respectively forbidden by the time-reversal symmetry and the inversion symmetry. The leading Hall response is of the third order.


We perform first-principles calculations based on the density functional theory (DFT), as implemented in the \texttt{Quantum ESPRESSO} package~\cite{giannozzi2009,giannozzi2017}.
The exchange-correlation functional is treated with the generalized gradient approximation using the Perdew-Burke-Ernzerhof (PBE) realization~\cite{perdew1996}.
The projector augmented wave pseudopotentials are adopted~\cite{blochl1994}.
We take 16 electrons for Fe (3$s^{2}$3$p^{6}$3$d^{6}$4$s^{2}$) and 16 electrons for Se (3$d^{10}$4$s^{2}$4$p^{4}$) as valence electrons.
Spin-orbit coupling effects are included in the calculation.
The kinetic energy cutoff for wave functions is set to 80 Ry, and the kinetic energy cutoff for charge density is fixed to 600 Ry.
For the self-consistent calculations, the Brillouin zone integration is performed on a Monkhorst-Pack grid mesh of $20 \times 20 \times 1$ $k$ points.
The energy convergence criterion is set to be $10^{-6}$ eV.
An experimental lattice parameter with $a=3.90$ \AA\ was used in our calculation~\cite{liu2012}.
To calculate band geometric quantities, a Wannier tight-binding Hamiltonian consisting of Fe-$3d$ and Se-$4p$ orbitals is constructed using the Wannier90 package~\cite{pizzi2020}.

The calculated electronic band structure is shown in Fig.~\ref{fig3}(c), along with the projected density of states (PDOS). One can see that the system is metallic. The low-energy states are mainly from Fe-$3d$ orbitals. The Fermi surface consists of an electron pocket at $M$ and a hole pocket at the $\Gamma$ point. These results are consistent with the previous works~\cite{liu2012,huang2017}.

\begin{figure}[tb!]
\centering
\includegraphics[width=1\columnwidth]{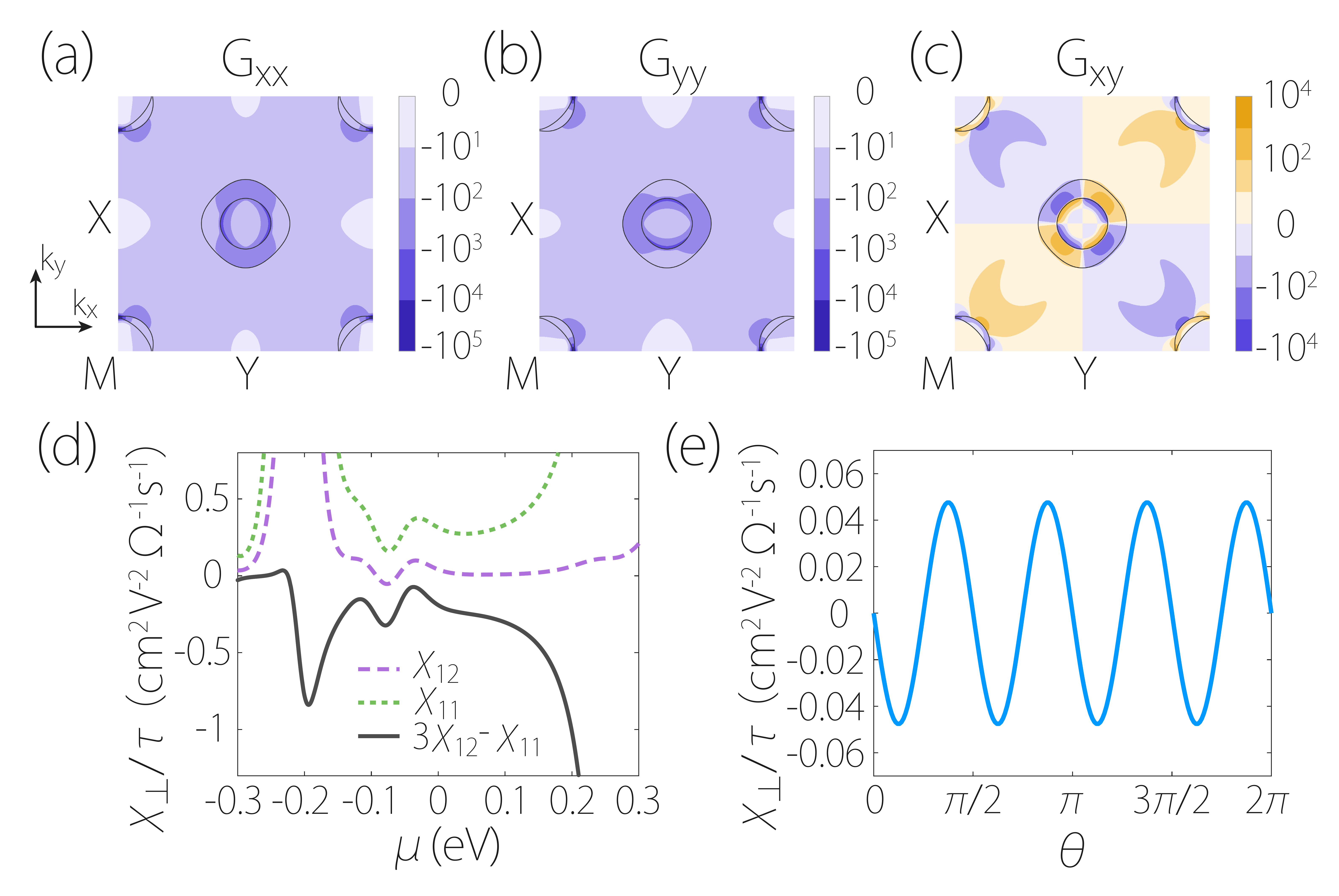}
\caption{ (a-c) Distribution of BCP tensor elements in the Brillouin zone for monolayer FeSe. Here, the units are \AA$^{2}\cdot$V$^{-1}$. The black lines depict the Fermi surface. (d) Variation of third-order conductivity tensor elements with the chemical potential. (e) Third-order transverse conductivity $\chi_\bot$ versus the angle $\theta$ ($\theta$ is the angle between the $E$ field and the [10] direction).
}
\label{fig4}
\end{figure}

Based on the band structure results, we further evaluate the BCP tensor for monolayer FeSe. In Fig.~\ref{fig4}(a-c), we plot the results of  $G_{ab}(\bm{{k}})$ for
all occupied states below the Fermi level. One observes that the tensor elements are mostly concentrated around the $\Gamma$ and $M$ points, because in these regions, there are band near-degeneracies slightly below the Fermi level that give the dominant contributions. $G_{xx}$ and $G_{yy}$ are connected with the $C_{4z}$ rotation. $G_{xy}$ manifests a quadrupole-like structure similar to that in the simple Dirac model. These are consistent with the crystal symmetry of the system.


Then, using the obtained BCP, the third-order conductivity tensor elements $\chi_{abcd}^{\text{I}}$ can be evaluated.
The crystal symmetry of monolayer FeSe ensures that there are eight
nonzero in-plane elements and four of them are independent, which are
$\chi_{xxxx}=\chi_{yyyy}$, $\chi_{xxyy}=\chi_{yyxx}$, $\chi_{xyxy}=\chi_{yxyx}$, and
$\chi_{xyyx}=\chi_{yxxy}$. Repeating the similar analysis as in the last section, for an applied in-plane electric field
$\bm{E}=E(\cos\theta,\sin\theta)$, the third-order transverse conductivity is obtained as
\begin{equation}
\chi_\bot(\theta)=-\frac{1}{4}(\chi_{11}-3\chi_{12})\sin4\theta,
\end{equation}
where $\theta$ is measured from the [10] direction, $\chi_{11}=\chi_{xxxx}$, and
$\chi_{12}=(\chi_{xxyy}+\chi_{xyxy}+\chi_{xyyx})/3$. The calculation results are shown in Fig.~\ref{fig4}(d). One observes that the $\chi$'s exhibit peak features at around $-0.2$ eV, which is mainly from the degeneracy point at $M$. At Fermi level, i.e., when $\mu=0$ eV, $\chi_\bot/\tau$
reaches a value about $-0.048$ cm$^{2}$V$^{-2}$$\Omega^{-1}$s$^{-1}$.
In the Drude theory, the longitudinal conductivity $\sigma= ne^{2}\tau/m^{*}$, with $n$ the carrier density and $m^*$ the effective mass,
the ratio $\chi_{\bot}/\sigma$ is independent of the scattering time. Estimated
from the experiments of monolayer FeSe~\cite{huang2021}, we take $n\sim10^{12}\ $cm$^{-2}$ and  $m^*\sim1.5m_{0}$ with $m_0$ the free electron mass. Then, we obtain {$\chi_{\bot}/\sigma\sim-1\times10^{-2}\ \mu\text{m}^{2}\text{V}^{-2}$}.
This value is much larger than the recent experimental result on multilayer WTe$_{2}$ ($\sim2\times10^{-4}\ \mu\text{m}^2\text{V}^{-2}$), so the effect should be observable in experiment. $\chi_{\bot}/\sigma$ will change when the chemical potential of monolayer FeSe is adjusted by doping or gating, and its magnitude can be enhanced at least by a factor of 4 when $\mu\sim -0.2$ eV.
Figure~\ref{fig4}(e) shows the angular dependence of $\chi_\bot$. Due to the additional mirror lines, the variation of $\chi_\bot$ with $\theta$ shows a period of $\pi/2$. The maximal values of $\chi_\bot$ are achieved when $\theta=\pm\frac{\pi}{8}+\frac{n\pi}{2}$ with $n\in\mathbb{Z}$. Such angular variation behavior can be directly tested in experiment with the angle-resolved measurement, as in Refs.~\cite{kang2019,lai2021}.

\section{Discussion and Conclusion}

Through this study, we now have a nice correspondence between the band geometric quantities and the Hall effects. The Berry curvature, Berry curvature dipole, and BCP play important roles in the first-, second-, and third-order Hall effects, respectively. We have demonstrated the possibility to theoretically evaluate the THE with first-principles calculations.
It also follows that the THE offers a way to probe the BCP.

The THE also provides a new characterization tool for materials with preserved time-reversal and inversion symmetries, which actually include a large class of materials. In these materials, both linear and second-order Hall effects are suppressed by symmetry, so the third order gives the leading order Hall response. It should also be noted that our theory is not limited to this class of materials. For materials with nonvanishing linear or second-order Hall response, the THE can still be detected in experiment by the lock-in technique and compared with our theory.

Here, we wish to comment on the limitation of this work and possible future directions. First, in this study, we have focused on the THE linear in the relaxation time, i.e., the $\chi^\text{I}$ term, which is closely connected with the BCP. In comparison, the $\chi^\text{II}$ term does not involve the BCP and it is proportional to $\tau^3$. {Hence, the relative importance of $\chi^\text{I}$ and $\chi^\text{II}$ would depend on extrinsic factors such as sample quality and measurement conditions. Nevertheless, this term can in principle also be measured in experiment and evaluated with first-principles calculations. Such studies shall be done in future works.}

Second, the derived formulas for the third-order conductivity also contains the longitudinal transport component. Of course, for longitudinal transport, the lower-order responses would also exist (the second-order one may be suppressed in centrosymmetric materials). Nevertheless, it should be possible to separate out the third-harmonic signal and compare with our theory.

Third, our theory here is within the simplest relaxation time approximation. There could exist additional contributions when going beyond this approximation. For example, the relaxation time may have variations for different states on the Fermi surface and may have corrections from the applied field. { Particularly, different relaxation processes (such as for momentum, spin, energy, etc.) typically have different relaxation times, and the difference may have important effects on nonlinear transport. For example, energy relaxation could be closely connected to $E^2$ terms.} In addition, more sophisticated treatment of the collision integral may result in contributions analogous to side jump and skew scattering in the linear anomalous Hall response~\cite{nagaosa2010}. { All these possible effects are beyond our current treatment. How good the relaxation time approximation is for describing nonlinear transport is still an open problem.
In this regard, our current theory has the advantage to be readily implemented in first-principles calculations for real materials. The obtained 
result can be directly compared with experiment, which would help us to assess the accuracy of the approach. We note that previous calculations within the relaxation time approximation for second-order response and our recent work on third-order response have found good agreement with experiments on 2D WTe$_2$ and MoTe$_2$~\cite{ma2019,kang2019,lai2021}. Nevertheless, more systematic studies are needed for future research, and the current work offers a good starting point for developing a more comprehensive theory for third-order transport effects.}

Finally, the study may be extended to magnetic materials. Since magnetism directly breaks the time-reversal symmetry, we expect there will be more terms contributing to the nonlinear conductivity. Particularly, terms with even powers of $\tau$ will be allowed by symmetry, and there should exist an ``intrinsic'' contribution that is independent of scattering. Such nonlinear responses in magnetic materials would be an interesting topic to explore in subsequent studies.

In conclusion, in this work, we have developed a theory for the third-order current transport based on the extended semiclassical formalism. We demonstrate that the BCP plays a key role in the third-order conductivity. We apply the theory to the 2D Dirac model and exhibit the important features of the BCP and the third-order current. We further demonstrate the combination of our theory with first-principles calculations in studying concrete material systems, which predicts a sizable effect in monolayer FeSe. Our work highlights the significance of the BCP, establishes the foundation for the study of third-order responses, and suggests the THE as a new characterization tool for a large class of materials.

\begin{acknowledgements}
The authors thank K.-Y. Lee and D. L. Deng for valuable discussions. This work is supported by the Singapore Ministry of Education AcRF Tier 2 (Grant No.~MOE2019-T2-1-001), the National Research Foundation CRP program (Grant No. NRF-CRP22-2019-0004), the National Natural Science Foundation of China (No. 11604273), and the UGC/RGC of Hong Kong SAR (AoE/P-701/20). We acknowledge computational support from the Texas Advanced Computing Center and the National Supercomputing Centre Singapore.
\end{acknowledgements}

\bibliography{3rdref}

\end{document}